\documentclass[10pt,letterpaper,twocolumn]{article}
\usepackage[english]{babel}
\usepackage{graphicx}

\newcommand{\alt}{\mathbin{\lower 3pt\hbox
   {$\rlap{\raise 5pt\hbox{$\char'074$}}\mathchar"7218$}}}
\newcommand{\agt}{\mathbin{\lower 3pt\hbox
   {$\rlap{\raise 5pt\hbox{$\char'076$}}\mathchar"7218$}}}

\begin{document}

\setcounter{footnote}{0}
\setcounter{equation}{0}
\setcounter{figure}{0}
\setcounter{table}{0}

\title{\large\bf Unusual phase transition in 1D
localization \\ and its
observability in optics }

\author{\small
S.\,I\,.Bozhevolnyi$^{(1)}$,\,\,  I.\,M.\,Suslov$^{(2)}$ \\
\small ${\phantom{gg}}^{(1)}$ Centre for Nano Optics, University
of Southern Denmark, Campusvej 55, \\
\small  DK-5230 Odense, Denmark \\
\small ${\phantom{gg}}^{(2)}$  P.L.Kapitza Institute for Physical
Problems,   \small 119334 Moscow, Russia \\
\small E-mail: suslov@kapitza.ras.ru\\
{} \\
\parbox{150mm}{\footnotesize \,Localization of
electrons in 1D disordered systems is usually described
in the random phase approximation, when
distributions of phases $\varphi$ and $\theta$, entering the
transfer matrix, are considered as uniform. In the general case,
the random phase approximation is violated, and the evolution
equations are written in terms of the Landauer resistance
$\rho$ and the combined phases $\psi=\theta-\varphi$
and $\chi=\theta+\varphi$. The distribution of the phase
$\psi$ is found to exhibit an unusual phase transition at the
point ${\cal E}_0$ when changing the electron energy ${\cal E}$,
which manifests itself
in the appearance of the imaginary part of $\psi$.
The distribution of resistance $P(\rho)$ has no singularity at
the point ${\cal E}_0$, and the transition seems
unobservable in the framework of
condensed matter physics. However, the
theory of 1D localization is immediately applicable to
the scattering of waves propagating in a single-mode optical waveguide.
Modern optical methods open a way to measure phases $\psi$ and $\chi$.
As a result, the indicated phase transition becomes observable.  } }

\date{}
\maketitle


\setcounter{footnote}{0}
\setcounter{equation}{0}
\setcounter{figure}{0}
\setcounter{table}{0}

\begin{center}
{\bf 1. Introduction}
\end{center}

Localization of electrons in 1D disordered systems
can be described by different methods \cite{9}. The most efficient
of them is based on the use of the transfer matrix
$T$, relating the amplitudes of plane waves on the left
($Ae^{ikx}+Be^{-ikx}$) and on the right ($Ce^{ikx}+De^{-ikx}$) of
a scatterer,
$$
\left ( \begin{array}{cc} A \\ B \end{array} \right)\,
=  T \left ( \begin{array}{cc} C \\ D \end{array}\right)
\,.
\eqno(1)
$$
The matrix $T$ can be parametrized in the form \cite{1}
 $$
 T= \left ( \begin{array}{cc} \!\!\! 1/t\! &\! - r/t \!\!\\
\!\!- r^*/t^* \!&\! 1/t^* \!\!\!\end{array} \right)\,
= \left ( \begin{array}{cc}
\!\!\sqrt{\rho\!+\!1}\, e^{i\varphi}\!\! &
\!\!\sqrt{\rho} \,e^{i\theta}\!\!
\\ \!\!\sqrt{\rho}\, e^{-i\theta}\!\!
&\!\! \sqrt{\rho\!+\!1}\,
e^{-i\varphi}\!\! \end{array} \right)\,,
\eqno(2)
$$
if the presence of the time-reversal invariance is suggested;
here $t$ and $r$ are the amplitudes of transmission
and reflection, while $\rho=|r/t|^2$ is the dimensionless
Landauer resistance \cite{2}.
If scatterers are arranged successively, their transfer
matrices are multiplied. The transfer matrix $T$ of a weak
scatterer is close to the unit one, allowing to derive
the differential evolution equations for its parameters.

Usually, such equations are derived in the random phase
approximation, which is successfully used in the different
areas of physics, though its limitations are also understood
\cite{100}--\cite{106}. Its application in respect to the
transfer matrix suggests the uniform distributions of $\varphi$
and $\theta$  \cite{3}--\cite{8}.  Such approximation is
sufficiently good for weak disorder well within the allowed band,
as is widely  accepted (see references in \cite{9,10,11}). The
fluctuation states in the forbidden band are considered
infrequently \cite{12,13,14} and only at the level of  wave
functions. A systematic analysis shows that the random phase
approximation is strongly violated near the initial band edge and
in the forbidden band of an ideal crystal \cite{15}. In the
general case, the evolution equations are written in terms of the
Landauer resistance $\rho$ and the combined phases (Sec.2)
$$
\psi=\theta-\varphi\,,\qquad
\chi=\theta+\varphi\,.
\eqno(3)
$$
The phase $\chi$ does not affect the evolution of $\rho$
and is not interesting for the condensed matter physics.
The distribution of the phase $\psi$
was considered in the papers \cite{16,17} and
undergoes the
peculiar phase transition at the point ${\cal E}_0$ under the
change of the electron energy  ${\cal E}$ \cite{17} (Sec.3).
Meanwhile, the distribution of resistance $P(\rho)$ has no
singularity at the point ${\cal E}_0$, and the transition  looks
unobservable in the framework of
condensed matter physics.

However, the approach developed previously \cite{15,16,17} is
immediately applicable to the scattering
of waves propagating in a
single-mode optical waveguide (Sec.4).
Existent optical methods (heterodyne approach, near-field
microscopy, etc.) are rather efficient, and
allows one to measure distributions of all parameters
$\rho$, $\psi$, $\chi$ inside the
waveguide\,\footnote{\,In this context,
the parameter $\rho$ does not
have a meaning of
Landauer resistance, but determines the amplitudes of
transmitted and reflected waves (Sec.5).} (Sec.6),
and the indicated phase transition appears to be
observable (Secs.5,\,6).
The general scheme of measurement
suitable for this purpose is described
in Sec.7.

\begin{center}
{\bf 2. Evolution equations }
\end{center}

\begin{figure}
\centerline{\includegraphics[width=3.4 in]{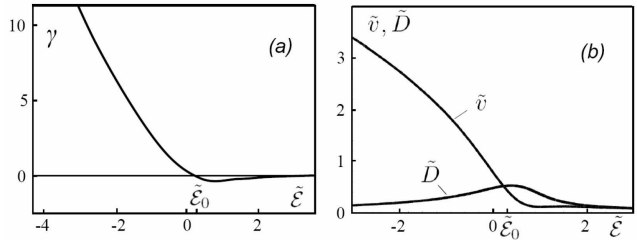}}
\caption{\small Dependence of parameters $\gamma$,
$\tilde v=v/W^{2/3}$ and $\tilde D=D/W^{2/3}$ on the reduced
energy $\tilde{\cal E}={\cal E}/W^{4/3}$, obtained from the
analysis of moments for the transfer matrix elements
\cite{15}. These moments are regular functions of energy,
which leads to regularity of the presented dependencies.
Smallness of $\gamma$ and the equality $v=D$, valid in the random
phase approximation, are realized only in the deep of the allowed
band. The point $\tilde{\cal E}_0$ corresponds to the phase
transition in the distribution $P(\psi)$.}
\label{fig1} \end{figure}

The most general evolution equation describes the change
of the mutual distribution $P(\rho,\psi,\chi)$ under
increasing of the system length $L$ and has the following
structure\,\footnote{\,Equation (4) was derived in the course of
preparation of the papers \cite{16,17}; its complete form was not
actual, and only its reduced variant (5) was published. Since the
phase $\chi$ becomes observable (Sec.6), derivation of (4) and
detailed analysis of the $\chi$ distribution will be a subject of
future publications.  Representation of the right-hand side as a
sum of full derivatives is a common property of the Fokker-Planck
equations, providing conservation of probability. The only
specific feature
of Eq.4 is independence on $\chi$ for  operators $\hat L$ and
$\hat M$. This property can be deduced from the fact, that the
evolution equation becomes closed on the level of two variables
$\rho$ and $\psi$.}
$$
\frac{\partial P}{\partial L}=
\left\{\vphantom{L^2} \hat L_{\rho,\psi} P \right\}'_\rho +
\left\{ \vphantom{L^2} \hat M_{\rho,\psi} P \right\}'_\psi \,
+\left\{ \vphantom{L^2} \hat K_{\rho,\psi,\chi} P \right\}'_\chi
\,,
\eqno(4)
$$
where $\hat K$, $\hat L$, $\hat M$ are operators, depending on
indicated variables.
The right-hand side is the sum of full derivatives, which
ensures the conservation of probability. Integration
of Eq.4 over $\chi$ allows to obtain the evolution
equation for $P(\rho,\psi)$
$$
\frac{\partial P}{\partial L}=
\left\{\vphantom{L^2} \hat L_{\rho,\psi} P \right\}'_\rho +
\left\{ \vphantom{L^2} \hat M_{\rho,\psi} P \right\}'_\psi \,,
\eqno(5)
$$
whose specific form is  given
in \cite{16,17}. In the large
$L$ limit, the typical values of $\rho$ are large, and the
operator $\hat M_{\rho,\psi}$ becomes independent of $\rho$;
then the solution of Eq.5 is factorized,
$P(\rho,\psi)=P(\rho) P(\psi)$, though a situation is
somewhat unusual for separation of variables \cite{17,18}.
The equation for $P(\psi)$ is split off, providing the
existence of the stationary distribution of the phase
$\psi$. The equation for $P(\rho)$ has a form
$$
\frac{\partial P(\rho)}{\partial L} =
D\,\frac{\partial}{\partial \rho}
\left[\,-\gamma(1\!+\!2\rho) P(\rho) +
\rho(1\!+\!\rho)\,\frac{\partial P(\rho)}{\partial \rho}
\,\right]   \,,
\eqno(6)
$$
and gives the log-normal distribution
$$
P(\rho)=\frac{1}{\rho \sqrt{4\pi D L}}
\exp\left\{-\frac{[\ln \rho-vL]^2}{4DL}\right\}\,,
\eqno(7)
$$
with $v=(2\gamma\!+\! 1)D$. The typical value of $\rho$
increases exponentially with  $L$, which
is an observable manifestation of 1D localization.
In the random phase approximation, the parameter $\gamma$
turns to zero, and equations (6), (7) coincide with
the previously obtained results
\cite{3}--\cite{8}. Dependencies of $\gamma$,
$D$, $v$ on the reduced energy $\tilde{\cal E}={\cal E}/W^{4/3}$
are shown in Fig.1, where ${\cal E}$ is the energy counted from
the initial band edge, and $W$ is an amplitude of the random
potential; all energies are measured in units of the hopping
integral of the 1D Anderson model, which is of the order of
the initial band width.
Strong violation of the random phase
approximation is thereby evident (Fig.1).

\begin{center}
{\bf 3. Phase transition in the distribution $P(\psi)$ }
\end{center}

The meaning of the phase transition in the $\psi$ distribution
consists in the fact that difference between the allowed
and forbidden bands survives (in a certain sense) in the
presence of the random potential, though a singularity in
the density of states is smoothed out. It resembles the
famous argumentation by Mott \cite{19}, that
the role of the allowed band edge
comes to the mobility
edge. Although the mobility edge is absent in the 1D case,
a general situation appears to be analogous. The point is that
the probe scatterer in the allowed band (${\cal E}>0$) is
described by the transfer matrix (2), while in the forbidden band
(${\cal E}<0$) it is described by the pseudo-transfer matrix
${\cal T}$ \cite{15}, relating coefficients of the increasing and
decreasing exponents on the left ($Ae^{\kappa x}+Be^{-\kappa x}$)
and on the right ($Ce^{\kappa x}+De^{-\kappa x}$) of the
scatterer.  In the simplest case, the matrix ${\cal T}$ is real
and corresponds to pure imaginary values of phases $\theta$ and
$\varphi$. Let us compare situations for ${\cal E}>0$ and
${\cal E}<0$: for a sufficient separation
in energy, the difference  between two types of
matrices can be made arbitrary large, and it cannot be
overcome
by inclusion
of a weak random potential. As a result, the
border-line between the true and
pseudo transfer matrices can only be
shifted, but not eliminated.
In practice, it is manifested
via the appearance of the imaginary part of the phase $\psi$
for energies ${\cal E}<{\cal E}_0$ \cite{17}.

The formal statements of the paper \cite{17} reduce to
the following. First of all, one should differ the
'external' and 'internal' phase distributions (Fig.2).
The internal phase distribution is realized in the deep of a
sufficiently long disordered system, and is independent of
boundary conditions. The phases entering the transfer matrix
refer to the 'external' phase distribution, which
depends on the boundary conditions and can be observed from the
side of the ideal leads. The internal phase distribution
continually transforms to the external one in the transient
region of the order of the localization length $\xi$, where
influence of interfaces is essential. Independence
of the limiting distribution $P(\rho)$ (Eq.7) from the
boundary conditions is evident from this picture, since it
is determined by the internal phase distribution.
However, the evolution equations contain not internal,
but the external phase distribution, and one wonders
why it does not affect the limiting distribution $P(\rho)$.
There is the second question, related with the first one, and it
can be formulated as follows: how the internal phase distribution
can be found, if it does not appear in the evolution equations?

\begin{figure}
\centerline{\includegraphics[width=3.4 in]{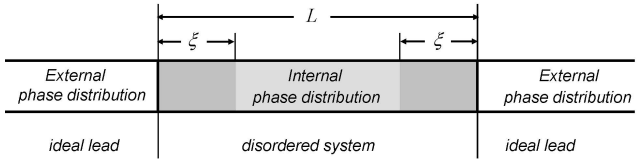}}
\caption{\small External and internal phase distributions.
}
\label{fig2}
\end{figure}

The above questions are resolved in the following
manner. It appears, that the phase $\psi$ is a 'bad' variable,
while the 'correct' variable should be defined as
$$
w=-{\rm cot}\,\psi/2 \,.
\eqno(8)
$$
The form of the stationary distribution $P(w)$ is
independent of boundary conditions, being determined by only
internal properties of the system.
If the boundary conditions are changed, it leads to three
effects: the scale transformation $w\to sw$ and
two translations $w \to w+w_0$ and $\psi\to\psi+\psi_0$.
The corresponding
changes of the distribution $P(\psi)$ are easily
predictable  \cite{17} and can be observed in the external
phase distribution. The evolution equations are invariant
in respect to translation $\psi\to\psi+\psi_0$, and the internal
phase distribution can be discussed under the fixed choice of
the origin. Invariance of the limiting distribution $P(\rho)$
under transformations $w\to sw$ and $w\to w+w_0$
is realized in the dynamical manner.
Analogously to
aperiodic oscillations of $P(\rho)$ \cite{15,22,22a},
in the region $L\alt \xi$
the scale factor $s$ and the translational shift $w_0$
undergo aperiodic oscillations as functions of $L$,
attenuating at large $L$. As a result, $s$ and $w_0$ tend
to the certain 'correct' values, which provide the
correct values of $D$ and $v$ in the limiting distribution
 (7). The indicated 'correct' values correspond to the internal
 phase distribution, and the latter can be found after return to
the variable $\psi$. Meanwhile, it appears that the
translational shift $w_0$ becomes complex-valued for
${\cal E}<{\cal E}_0$, indicating  the occurrence
of the imaginary part of $\psi$. This qualitative change
indicates existence of the unusual phase transition.

The point ${\cal E}_0$ is not singular for the Landauer
resistance  $\rho$, and the whole distribution $P(\rho)$
varies in its vicinity in a smooth manner  (Fig.1,b). As a
result, the described phase transition looks
unobservable in the framework of the condensed matter physics.
Fortunately, it has the observable manifestations in
optics  (Secs.5,\,6) in the form of the square
root singularities.

\begin{center}
{\bf 4. Analogy with optics }
\end{center}

Localization of classical waves was discussed in a number of
papers  \cite{200}--\cite{206}, \cite{10,11}. It includes
consideration of  weak \cite{201} and  strong \cite{202,203}
localization, absorption near a photon mobility edge \cite{200},
near-field mapping of intensity of optical modes in
disordered waveguides \cite{205},
and many other  aspects (see the review article \cite{204}.
The transfer matrix approach to the problem was discussed in
\cite{10,11,206}. In application to optics the corresponding
analysis reduces to a set of simple relations.

Propagation of electromagnetic waves in homogeneous
dielectric media is described by the wave equation
$$
c^2 \Delta \Psi-n^2 \frac{\partial^2 \Psi}{\partial t^2}=0\,,
\eqno(9)
$$
where $\Psi$ is any component of
the electric or magnetic field. If a medium
is spatially inhomogeneous, the refractive
index $n$ is varying along the coordinate $x$,
$$
n^2(x)=n_0^2+\delta
n^2(x)\,, \eqno(10)
$$
and for the monochromatic wave $\Psi\sim e^{i\omega t}$,
the wave equation can be written in the form
$$
\tilde c^2 \Delta \Psi+\left[\omega^2+\omega^2
\frac{\delta n^2(x)}{n_0^2}
 \right] \Psi=0 \,,\quad \tilde c=c/n_0\,.
 \eqno(11)
 $$
The latter exhibits
the same structure, as the Schr${\rm\ddot o}$dinger equation
for an electron with energy ${\cal E}$ and mass $m$ in the random
potential $V(x)$.  One can easily establish
the correspondence
$$
{\cal E}\,\Longleftrightarrow\, \omega^2 \,,\quad
\frac{1}{2m}\,\Longleftrightarrow\, \tilde c^2 \,,\quad
V(x)\,\Longleftrightarrow \,-
\omega^2 \frac{\delta n^2(x)}{n_0^2} \,.
 \eqno(12)
 $$
A certain difference from the condensed matter physics is
related to the $\omega$ dependence of the potential $V(x)$,
which of little importance,
if one is restricted by a small frequency interval of the
continuous spectrum.

\begin{figure}
\centerline{\includegraphics[width=3.2 in]{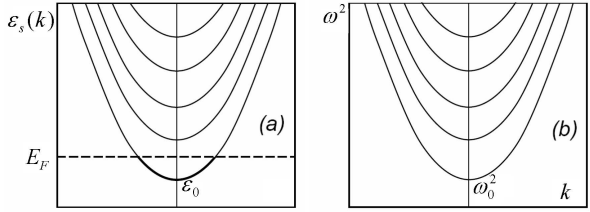}} \caption{
\small The spectrum of electrons in the metallic wire (a),
and the spectrum of waves propagating in a waveguide (b).  }
\label{fig3}
\end{figure}

The spectrum of waves propagating in a waveguide is
analogous to the spectrum of electrons in a metallic
wire. In the latter case, the transverse motion is quantized,
leading to a set of the discrete levels $\epsilon_s$. If
the longitudinal motion is taken into account, these levels
transform to one-dimensional bands with the dispersion
law (Fig.3,a)
$$
\epsilon_s(k)=\epsilon_s + k^2/2m\,.
 \eqno(13)
$$
To obtain a strictly 1D system, one should have a
sufficiently small Fermi level so that only the lowest band
is occupied.
In the presence of impurities, the lower boundary $\epsilon_0$
of the spectrum is smeared out due to the occurrence
of fluctuation
states for  ${\cal E}<\epsilon_0$. The dependencies shown in Fig.1
correspond to the energy ${\cal E}$ counted from $\epsilon_0$.

Analogously, quantization of the transverse motion in
a waveguide leads to
a set of discrete frequencies
$\omega_s=\tilde c  \kappa_s$, where $-\kappa_s^2$  are
eigenvalues of the 2D Laplace operator with the appropriate
boundary conditions \cite{21}.
The zero eigenvalue
is possible only in the case, when
the waveguide cross-section is multiply connected
(e.g. as in a coaxial cable).  For a singly connected
cross-section, the
minimum eigenvalue  $\omega_0$ is finite \cite{21}. If
the longitudinal motion is taken into account, the  following
branches of the spectrum are obtained
(Fig.3,b)\,\footnote{\,We have in mind a metal-coated
dielectric waveguide.
The coating thickness
should be of the order of the skin depth
in order to allow for partial field penetration
outside the waveguide  (see Sec.7). In the absence of
coating (a pure dielectric waveguide), one should take
into account
the frequency dependence of the effective potential $V(x)$
(see Eq.11),
which leads to lower restrictions for the allowed
values of the longitudinal  momenta $k$
(conditions for the total internal reflection are violated).
In addition, the parameters $\kappa_s$
cease to be constant and become $\omega$-dependent,
resulting in deviations from the
parabolic dependencies in Fig.3,b.}
$$
\omega^2_s(k)=\omega_s^2 + \tilde c^2 k^2\,.
 \eqno(14)
$$
To realize
a single-mode waveguide regime, one should operate
near the lower boundary $\omega_0$ of the spectrum. In the
presence of disorder, the spectrum boundary $\omega_0$
is smeared out due to the occurrence
of the fluctuation states.
Overall, the effects appearing in the electron system under the
change of the Fermi level can be observed in a single-mode
waveguide under the change of frequency $\omega$ in the vicinity
of $\omega_0$.

\begin{center}
{\bf 5. Detection of the phase transition }
\end{center}

Let a wave of the unit amplitude be incident
from the left side of a single-mode waveguide,
and comes through it with the amplitude $t$, being
reflected with the amplitude $r$. If there are point scatterers
in the waveguide, then a partial reflection occurs at any of
them. Thus, at an arbitrary point $x$ of the waveguide one
finds a superposition of two waves, propagating in opposite
directions.
The electric field $E(x,t)$
is determined by the  real part of this superposition, i.e.
$$
E(x,t)= {\rm Re}
\left[Ae^{ikx+i\omega t}+Be^{-ikx+i\omega t} \right] \,.
\eqno(15)
$$
With the transfer matrix $T$ being defined by Eqs.(1,2),
the amplitudes of the transmitted and reflected waves
are determined by the expression
$$
\left ( \begin{array}{cc} A \\ B \end{array} \right)\,=
T\,\left ( \begin{array}{cc} t \\ 0 \end{array}\right)=
\left ( \begin{array}{cc} |t|\sqrt{\rho\!+\!1}
\, e^{i\varphi-i\varphi_0} \\
|t|\sqrt{\rho} \, e^{-i\theta-i\varphi_0}
\end{array}\right) \,,
\eqno(16)
$$
where $\rho$, $\varphi$, $\theta$ are $x$-dependent and
$t=|t|e^{-i\varphi_0}$ is adopted.
If the amplitude $|t|$ is sufficiently small, then
the quantity $\rho$ is large in the whole waveguide,
except the vicinity of its right end.
Then $|A|\approx |B|$, and Eq.15
gives in this approximation
$$
E(x,t)=
{\rm Re} \left[|A|\,e^{ikx+i\varphi-i\varphi_0}
+|B|\,e^{-ikx-i\theta-i\varphi_0} \right]\,e^{i\omega t}
$$
$$   \approx
2|A|\cos{\left(kx+\chi/2\right)}
\cos{\left(\omega t-\psi/2-\varphi_0\right)} \,,
\eqno(17)
$$
where the combined phases $\psi$ and $\chi$ are defined
in Eq.3: they remain constant between the scatterers, and
change abruptly when passing through a scatterer. If
the concentration of impurities is large, then $\psi$ and
$\chi$ change with $x$ practically continuously, having
random variations on the
scale of the scattering length.

Since the field $E(x,t)$ can
be measured in principle,
both phases $\chi$ and $\psi$ are theoretically observable.
This is
the fundamental difference from the condensed matter
physics, where a superposition of waves refers to a wave
function, and should be squared in modulus to obtain the
observable quantities: in this case the phase $\psi$
is unobservable in principle. This phase would
become unobservable in optics, if only the average
intensity could be measured
(it means that equation (17) is squared and averaged over time).
It is easy to verify, that this conclusion remains valid also
for $|A|\ne |B|$.

\begin{figure}
\centerline{\includegraphics[width=3.1 in]{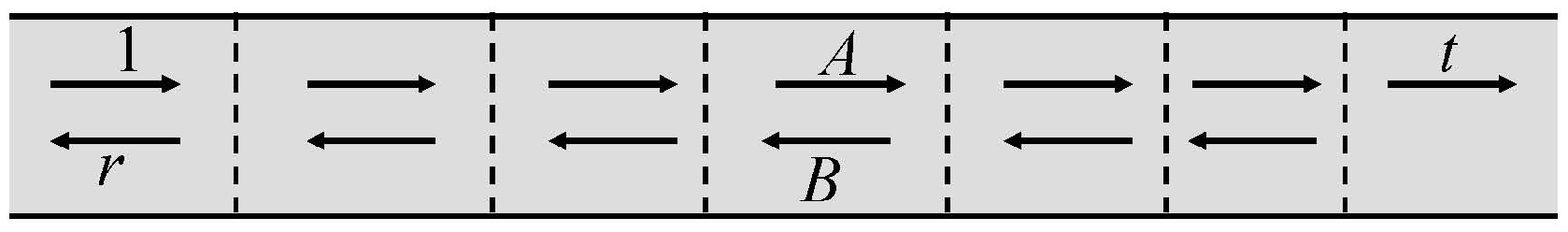}}
\caption{\small Propagation of waves in the
single-mode waveguide with point scatterers.  }
\label{fig4}
\end{figure}

Nevertheless, the occurrence
of the imaginary part of $\psi$
can be registered even in this case. If we suppose that
$$
\varphi=\varphi'+i\varphi''\,,\qquad
\theta=\theta'+i\theta''\,,
\eqno(18)
$$
then the amplitudes in the linear combination (15)
accept the form
$$
|A|= |t|\sqrt{\rho\!+\!1}
\, e^{-\varphi''} \,,\qquad
|B|=|t|\sqrt{\rho} \, e^{\theta''} \,.
\eqno(19)
$$
The flux conservation requires\,\footnote{\,A scattering is
considered as pure elastic. Inevitable Ohmic losses  in the
metal coating (see Footnote 2) are suggested to be
essentially weaker in comparison with localization
effects. } that
the condition $|A|^2=|B|^2+|t|^2$ be fulfilled at the
arbitrary point of the waveguide, which leads to relation
$$
(\rho\!+\!1)\, e^{-\varphi''} =
\rho \, e^{\theta''} +1
\eqno(20)
$$
giving $\theta''=-\varphi''$ for large $\rho$.
The imaginary part is absent in the phase $\chi$, but is
admissible for the phase $\psi$; in the latter case
$\psi''=2\theta''=-2\varphi''$, and in particular
$$
|A|= |t|\sqrt{\rho\!+\!1}
\, e^{\psi''/2} \,.
\eqno(21)
$$

According to \cite{17}, the imaginary part of $\psi$ appears
as a result of solution of certain equations, and its behavior
can be established from the general considerations. Let we
have the equation $F(x)=0$, where the function  $F(x)$
depends regularly on the external parameter $\epsilon$.
If at the point $\epsilon=0$ two real roots become
complex-valued, then the multiple root $x=p$ takes place
for $\epsilon=0$, and in its vicinity one has the equation
$$
(x-p)^2-\epsilon=0\,,
\eqno(22)
$$
which gives roots $p\pm \sqrt{\epsilon}$ for $\epsilon>0$
and roots $p\pm i \sqrt{|\epsilon|}$ for $\epsilon<0$.
Thereby, the appearance of the imaginary part is related
with a square root singularity. If the imaginary part of
$\psi$ appears for $\omega<\omega_c$, then it has a behavior
$$
\psi''\sim \sqrt{\omega_c-\omega}\,\Theta(\omega_c-\omega)\,.
\eqno(23)
$$
According to  \cite{17},  the distribution $P(\rho)$ is not
singular at the point $\omega_c$ (Fig.1,b). It  refers to
a value of $\rho$ at the arbitrary point of the waveguide, and
in particular to its value at the whole length $L$,
which is related with $t$ as $|t|=(1+\rho)^{-1/2}$. Therefore,
the singularity in the amplitude (21) is completely
determined by the quantity $\psi''$ and has a square root
character.

\begin{figure}
\centerline{\includegraphics[width=3.3 in]{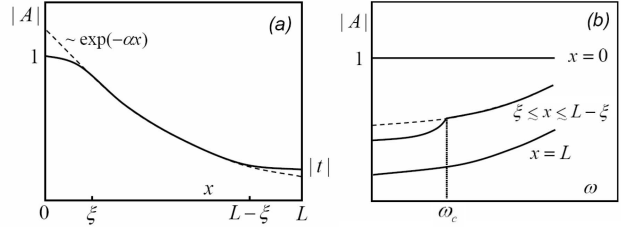}}
\caption{\small  (a) Dependence of the amplitude $|A|$
of the transmitted wave on the coordinate $x$ inside the
waveguide.  (b) The amplitude $|A|$ versus a frequency $\omega$
in the vicinity of the phase transition.  }
\label{fig5}
\end{figure}

The general picture looks as follows (Fig.5). In whole, the
modulus of $A$ changes in the waveguide according to the
exponential law, $|A|\sim e^{-\alpha x}$, but deviations from it
arise on the scale $\xi$ near the ends  due to influence of
boundary conditions (Fig.5,a): in particular, $|A|=1$ for $x=0$
and $|A|=|t|$ for $x=L$. The latter quantity is related with
$\rho$ and is a regular function of $\omega$.  However, in the
deep of the waveguide the amplitude $|A|$ has a square root
singularity (Fig.5,b), which can be registered already in
 the measurements of the average intensity.

According to \cite{17}, the critical point ${\cal E}_0$
is situated in the allowed band at the distance of order
$W^{4/3}$ from the band edge. Correspondingly, in optics
the critical point $\omega_c$
is greater
than the boundary frequency $\omega_0$, while a distance between
them is determined by the degree of disorder.

\begin{center}
{\bf 6. Observability of phases $\psi$ and $\chi$
} \end{center}

Measurements of the time dependence
at optical frequencies are usually impossible. However,
observability of the phase $\psi$ can be
provided with heterodyne technique, in which
the measured electric field $E(x,t)$ is mixed with
the additional field $E_s(x,t)$, whose
frequency is shifted by a small quantity $\Omega$:
$$
 E+E_s= {\rm Re} \left\{|E|e^{i\omega
t+i\varphi_E}+ |E_s| e^{i(\omega+\Omega) t+i\varphi_s} \right\}
\,.  \eqno(24)
$$
Considering the intensity averaged
over fast time oscillations, one has
$$
2\overline{(E+E_s)^2}=|E|^2+|E_s|^2+2|E||E_s|\cos{\left(\Omega
t+\varphi_s-\varphi_E\right)} \,.
\eqno(25)
$$
Substituting $E(x,t)$, corresponding to expression (17),
one obtains
$$
2\overline{(E+E_s)^2}=\left\{4|A|^2\cos^2{\left(kx+\chi/2\right)}
+|E_s|^2 \right\}+
$$
$$
+2|A|\cos{\left(kx+\chi/2\right)}\cdot
2|E_s|\cos{\left(\Omega t+\psi/2+\varphi_0+\varphi_s\right)}\,.
\eqno(26)
$$
so both phases $\chi$ and $\psi$ are observable, and
can be extracted from the experiment
by the following treatment.

The stationary first term and the oscillatory second term in
Eq.(26) can be separated
by the Fourier filtering in the time domain.
The constant term $|E_s|^2$
can then be easily extracted,
since the smallest
value of the first term in
the braces is zero.
Since the cosine changes regularly and reverses
sign at any zero, the square root from the first
term in braces can be extracted
to inessential common sign.
As a result, two combinations
would separately become known
$$
|A|\cos{\left(kx+\chi/2\right)} \quad \mbox{\rm and} \quad
|E_s|\cos{\left(\Omega t+\psi/2+\varphi_0+\varphi_s\right)} \,.
\eqno(27)
$$
The factor $|E_s|$ in the second combination is determined by
the amplitude of its
temporal oscillations\,\footnote{\,Another
way to reach the same result is to make measurements
for several values of $|E_s|$ and fit the right-hand side of (26)
by the dependence $\alpha +\beta |E_s| +\gamma |E_s|^2$.},
while its $x$ dependence can be
attributed to the spatial dependence of the phase  $\psi$.

The treatment of the first combination (27) is
complicated by the fact that the amplitude $|A(x)|$ does not
follow strictly
the exponential dependence $\exp(-\alpha x)$, but exhibits
significant fluctuations around it according to the
log-normal distribution (7). The
appropriate treatment
looks as follows:

1. Find a value of $k$ by
evaluating the average spatial period of oscillations.

2. Find values of $\chi$ at the sequence of discrete
points, which are maxima, minima and zeroes of the oscillating
dependence, by assessing deviations of their position
from those of the purely cosine function. If the value
of $k$ is estimated correctly, then the obtained $\chi$ values
would fluctuate around a constant value
and not exhibit a systematic growth.
As a result, one obtains a
sufficient material for the analysis of the $\chi$ distribution.

3. Find values of $|A(x)|$ at the points of maxima and
minima. These values would provide sufficient data
for verifying the log-normal distribution, and allow
to reveal
systematic deviations from the exponential dependence
near the waveguide ends.

Observability of the phase $\psi$ provides additional
possibilities for registration of the phase transition.
If one introduce the variable $w$ defined in Eq.(8),
then the moments of the distribution $P(w)$ (e. g.
$\langle w \rangle$) will have the singularities
$\sqrt{\omega-\omega_c}$ in the region $\omega>\omega_c$. The
phase $\chi$ does not affect the evolution of $P(\rho)$ and
was not studied in the papers \cite{16,17}. However,
the possibility of its observation
in optics makes such studies
to be actual.

\begin{center}
{\bf 7. The general measurement scheme }
\end{center}

\begin{figure}
\centerline{\includegraphics[width=3.1 in]{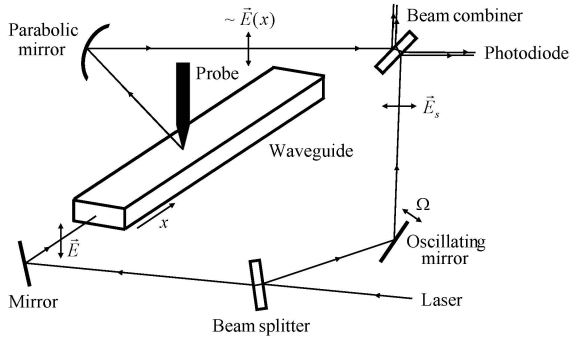}} \caption{
\small Measurement of the electric field in the
waveguide using the scanning near-field optical
microscope.  }
\label{fig6}
\end{figure}

The electrical field in a waveguide can
be measured
using methods of the scanning near-field optical microscopy
\cite{23,24,25}. In the scheme considered
below, a near-field microscope probe
(which is usually a fragment of an optical fiber with
a pointed tip) is used
not for the immediate field detection,
but only as a source of scattering\footnote{\,It can
be replaced by a needle of a scanning tunneling microscope,
which in the presence of metal coating
(see  Footnote 2) allows one to use all
advantages of the scanning tunneling electron microscopy.  }
with subsequent use of a remote detector.
A wave propagating in the waveguide penetrates beyond its
boundaries due to the tunneling effect
and can be scattered by a probe tip
located close to a waveguide surface.  For subwavelength-sized probe tips,
the scattering occurs in the Rayleigh
regime with the field of the scattered wave being
proportional to the local electric field $E(x,t)$
at the point of scattering $x$.\footnote{\,In the Rayleigh
scattering, the electromagnetic field of the scattered wave is
determined (in the main approximation) by the electric field of
the incident wave and does not depend on the wave vector of the
latter \cite{21}.  As a result, two waves entering the
superposition (15) are scattered equally, and the total field of
the scattered wave appears to be proportional to the electric
field in the waveguide.}

The general scheme of field
measurement
looks as follows (Fig.6).
A laser beam is split into two parts, one of which is
directed
into a waveguide and eventually scattered by a
microscope probe tip.
The scattered light is collected by a
parabolic mirror and directed to a
beam combiner. The second part of the laser beam is
reflected by an oscillating mirror, acquiring
a small frequency shift $\Omega$ due to the Doppler effect. Since
the mirror velocity is variable, it leads to a variable shift
$\Omega$.  This problem can however be solved, if the time
dependence is registered at the discrete points, equally spaced
by the period of mirror oscillations. Another possibility
consists in realization of the saw-toothed regime of
oscillations instead of the harmonic one.
After the mirror, the beam is directed to the beam combiner,
where it is mixed with the first beam and
follows to a photodiode
for measurement of intensity.

The described scheme was realized in studies reported in the
paper \cite{26}, where additional experimental details can be
found. In contrast to the papers \cite{207,208}, where only the
transmission matrix was measured, Ref.\cite{26} presents the
experimental approach allowing to measure the phase
distribution inside the waveguide.  
Ref.\cite{26} deals with surface plasmon polaritons, i.e. 
electromagnetic excitations, propagating along a metal-dielectric 
interface. In the transverse direction they are localized on the 
scale about 100nm, which is an order of magnitude less than a typical 
size of a single-mode
waveguide. The latter is not essential, since their transverse
profile extends to the air side in the same manner, as for waves
propagating in a waveguide and tunneling through its boundaries.
Correspondingly, they are subjected to analogous scattering by
the microscope probe. The spectrum of surface waves is analogous
the that of the metallic waveguide (Fig.3,b), and the single-mode
regime was realized in Ref.\cite{26}. However, the measurements of
Ref.\cite{26} were not concerned with light propagation in
disordered systems, but only with characterization of regular
modes in homogeneous waveguides.

Essentially new measurements are necessary for testing the
validity of claims made in the present paper. These
experiments  should overcome certain experimental difficulties.
 Firstly, the most promising waveguide configuration
 should be established,  along with the approach for introducing
 properly disordered defects into the waveguide. Extensive
analysis is necessary to find the parameter range, where
the localization effects will be dominating over the light
absorption inside the waveguide and radiative
losses through its boundaries.
Finally,  the actual experiments should be executed using a
tunable laser allowing to change the light frequency, and
its tunability range should cover the targeted
phase transition. This is a very ambitious and demanding project
going far beyond the framework of the current work.

\begin{center}
{\bf 8. Conclusion }
\end{center}

It has been shown that all results obtained for electrons
in 1D disordered
systems are immediately applicable to the propagation
of electromagnetic waves in single-mode optical waveguides.
It has also been argued that modern optical techniques,
such as near-field microscopy,
enable  measurements of all parameters entering the transfer
matrix. In particular, it becomes possible to observe the phase
transition for the distribution $P(\psi)$, which seems
unobservable in the framework of the condensed matter physics.
We believe that the results obtained might stimulate
the corresponding experimental activities that would, in turn,
shed more light on intricate effects in both optical and
electron localization phenomena.

\end{document}